\documentclass[twocolumn,showpacs,preprintnumbers,aps,pra,floatfix,superscriptaddress]{revtex4-2}

\usepackage{amssymb}
\usepackage{amsmath}

\usepackage{color}
\usepackage{graphicx}      
\usepackage{subfigure}
\usepackage{dcolumn}
\usepackage{textcomp}
\usepackage{longtable}     
\usepackage{url}           
\usepackage{bm}            
\usepackage{appendix}
\usepackage{hyperref}
\usepackage{soul}
\usepackage{physics}

\newcommand{\NaRb}{$^{23}$Na$^{87}$Rb}
\newcommand{\Xstate}{$X^1\Sigma^+$}

\begin{document}

\title{Characterization of the lowest excited-state ro-vibrational level of \NaRb}

\author{Junyu He}
\affiliation{Department of Physics, The Chinese University of Hong Kong, Shatin, Hong Kong SAR, China}
\author{Junyu Lin}
\affiliation{Department of Physics, The Chinese University of Hong Kong, Shatin, Hong Kong SAR, China}
\author{Romain Vexiau}
\affiliation{Universit{\'e} Paris-Saclay, CNRS, Laboratoire Aim{\'e} Cotton, 91405 Orsay, France}
\author{Nadia Bouloufa}
\affiliation{Universit{\'e} Paris-Saclay, CNRS, Laboratoire Aim{\'e} Cotton, 91405 Orsay, France}
\author{Olivier Dulieu}
\affiliation{Universit{\'e} Paris-Saclay, CNRS, Laboratoire Aim{\'e} Cotton, 91405 Orsay, France}
\author{Dajun Wang}
\email{djwang@cuhk.edu.hk}
\affiliation{Department of Physics, The Chinese University of Hong Kong, Shatin, Hong Kong SAR, China}
\date{\today}

\begin{abstract}

Starting from an ultracold sample of ground-state \NaRb\,molecules, we investigate the lowest ro-vibrational level of the $b^3\Pi$ state with high resolution laser spectroscopy. This electronic spin-forbidden $X^1\Sigma^+ \leftrightarrow b^3\Pi$ transition features a nearly diagonal Franck-Condon factor and has been proposed useful for probing and manipulating the ultracold molecular gas. We measure the transition strength directly by probing the ac Stark shift induced by near resonance light and determine the total excited-state spontaneous emission rate by observing the loss of molecules. From the extracted branching ratio and the theoretical modeling, we find that the leakage to the continuum of the $a^3\Sigma^+$ state plays the dominant role in the total transition linewidth. Based on these results, we show that it is feasible to create optical trapping potentials for maximizing the rotational coherence with laser light tuned to near this transition.           



\end{abstract}

\pacs{}

\maketitle

\section{Introduction}
\label{sec:introduction}
In recent years, the research direction of ultracold polar molecules (UPMs) has received an increasingly intensive attention~\cite{Bohn2017}. Much of this interest stems from the permanent electric dipole moment of polar molecules which, in ultracold temperatures, can be harnessed for a broad range of applications in quantum simulation and quantum information processing~\cite{Yan:2013,Trefzger.2011,Baranov.12,Demille.02}. UPMs are also natural candidates for investigating chemical reactions at ultra-low energies where quantum mechanical effects have been observed to play a dominated role~\cite{Ni.08,Ye.18,Gregory:2019,Hu19,Hu:2020}. 

Another great asset of polar molecules is their rich internal structures, including various electronic, vibrational, rotational and nuclear spin states. For UPMs in the electronic ground state, vibrational levels have been used to control the two-body chemical reactivity~\cite{Ye.18}, rotational levels can serve as building blocks of qubits with dipolar interactions when coupled by microwave~\cite{ni.18,Sawant2020,Hughes2020}, while nuclear spin levels have long coherence time and are nice candidate for quantum information storage~\cite{Park:2017aa}.            

On the other hand, the internal structures of polar molecules also pose a great challenge for cooling them to ultracold temperatures since, generally speaking, finding a cycling transitions necessary for efficient laser cooling becomes more difficult. Currently, the most successful method for creating high phase-space density samples of UPMs is by associating ultracold heteronuclear alkali atoms followed by a stimulated Raman process to transfer the resultant weakly-bound molecules to the ground state~\cite{Ni.08,Takekoshi14,Molony14,Park15,Guo.16,Kai20}. Nevertheless, for some special ground-state molecules with diagonal Franck-Condon factors (FCFs) and thus quasi-cycling transitions, direct laser cooling to $\mu$K temperatures has become possible \cite{Ding.20,Anderegg.17,Collopy.18,Truppe:2017,Barry:2014}.

\begin{figure}[t]
  \includegraphics[width=0.9\linewidth]{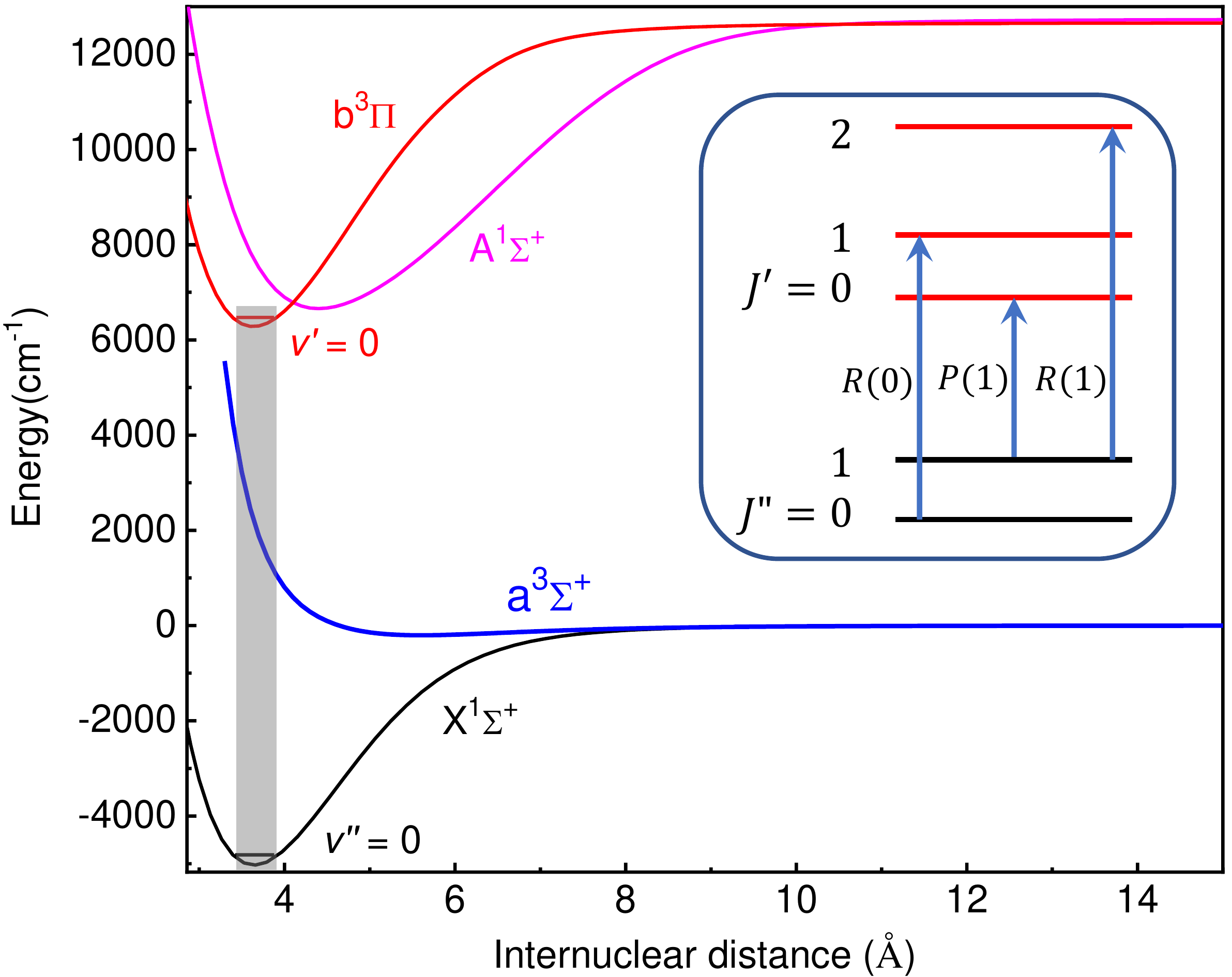}
\caption{Probing the ($b^3\Pi, v' = 0$) level of $^{23}$Na$^{87}$Rb with an ultracold sample prepared in the ($X^1\Sigma^+, v'' = 0$) ground state. The $v''=0 \leftrightarrow v'$ transition features a diagonal FCF (vertical gray bar). Inset: Rotational branches accessed in this work.  }
\label{fig1}
\end{figure}

For bi-alkali polar molecules, such diagonal FCFs can also be found, for instance between the $v = 0$ vibrational level of the $X^1\Sigma^+$ state and the $v' = 0$ vibrational level of the first excited triplet state $b^3\Pi$. As shown in Fig.~\ref{fig1}, this electronic spin-forbidden transition is made weakly allowed by the $b^3\Pi-A^1\Sigma^+$ mixing in the excited state. This transition has been investigated in several bi-alkali polar molecules. For KRb, narrow-line laser cooling with this transition was proposed~\cite{Kobayashi.14}. For NaK, engineering rotational-state dependent trapping potential with low photon scattering rate has been demonstrated~\cite{See.18}. For RbCs, such ``magic'' trapping potential has been extended to multiple rotational levels~\cite{guan.21}.

In this work, we investigate the $(X^1\Sigma^+, v'' = 0, J'') \leftrightarrow (b^3\Pi, v' = 0, J')$ electronic spin-forbidden transition of the \NaRb\, molecule which has a diagonal FCF. Starting from ultracold ground-state \NaRb\, molecules prepared in $J'' = 0$ and 1 rotational levels, we observe all the $P$ and $R$ branches and the accompanying nuclear spin structures. With the rotationally closed $J'' = 1 \leftrightarrow J' = 0$ [the $P(1)$] transition, we calibrate the transition dipole moment (TDM) and measure a total linewidth of $2\pi \times 225(5)$~kHz. Combined with results from quantum chemistry calculations, we find that the total linewidth is dominated by spontaneous decay to the continuum of the $a^3\Sigma^+$ state. 

The rest of this paper is organized as follows. In Sec.\ref{sec:Theoretical} we describe some basic properties of the $(b^3\Pi, v' = 0)$ level. We then describe our experimental apparatus and procedure in Sec.\ref{sec:Experimental}. In Sec.\ref{sec:Results}, we present the $(b^3\Pi, v' = 0)$ spectrum and the calibration of the $(X^1\Sigma^+, v'' = 0, J'' =1) \leftrightarrow (b^3\Pi, v' = 0, J' = 0)$ transition strength. We then discuss some possible applications of this transition in Sec.\ref{sec:Discussion}, and conclude the paper in Sec.\ref{sec:Conclusion}.   

\section{The $(X^1\Sigma^+, v = 0) \leftrightarrow (b^3\Pi, v' = 0)$ transition}
\label{sec:Theoretical}

The excited-state $b^3\Pi-A^1\Sigma^+$ complex in \NaRb\, has been investigated in detail both theoretically and experimentally in previous works~\cite{Docenko.2007,Guo.2017}. However, to our knowledge, the lowest ro-vibrational level of the $b^3\Pi$ state has not been observed. As indicated by the vertical gray bar in Fig.~\ref{fig1}, the bottom of the $b^3\Pi$ potential matches with that of the $X^1\Sigma^+$ potential very well which is a clear hallmark of the existence of diagonal FCFs. Indeed, our calculation based on the RKR potentials~\cite{Docenko.2007} gives a $v'' = 0 \leftrightarrow v' = 0$ FCF of 96.9\%, similar to those found in molecules suitable for direct laser cooling.  

The triplet-singlet mixing in the $b^3\Pi-A^1\Sigma^+$ complex is caused by spin-orbit coupling (SOC) which splits the $b^3\Pi$ state into components $b_{0^\pm}$, $b_1$ and $b_2$ with total electronic angular momentum ${\Omega} = 0^\pm, 1$ and 2, respectively. The SOC effect is the strongest near the crossing of the $b^3\Pi$ and $A^1\Sigma^+$ potentials at around $4.1a_0$. As shown in Fig.~\ref{fig1}, the bottom of the $b^3\Pi$ state actually lies below this crossing and is thus less perturbed by the $A^1\Sigma^+$ state. Indeed, from our modeling~\cite{Guo.2017}, the several low lying vibrational levels of the $b^3\Pi$ state all have $>99\%$ triplet character. The $v' = 0$ level, which is of interest here, contains \textcolor{red}{99.75}\% $b_{0^+}$ component and only \textcolor{red}{0.25}\% $A^1\Sigma^+$ component. Nevertheless, it is this small $A^1\Sigma^+$ component coupled with the large FCF which makes the $v'' = 0 \leftrightarrow v' = 0$ transition possible.   

Including the SOC in the ground state, the bottom of $X^1\Sigma^+$ is of pure $\Omega = 0^+$. The $v'' = 0 \leftrightarrow v' = 0$ transition is thus mainly of $\Omega = 0 \leftrightarrow \Omega = 0$ character. The rotational selection rules only allow the $P$ ($R$) branches with $\Delta J = J' - J'' = -1~(+1)$ to occur while the $Q$ branch with $\Delta J = 0$ is forbidden. Starting from $J'' = 0$ and 1 in the ground state, only $J' = 0$, 1 and 2 can be reached in the excited state (inset of Fig.~\ref{fig1}). As $J' = 0$ can only decay to $J'' = 1$, the $P(1)$ branch is rotationally closed. This transition is the focus of the current work. This rotational selection rule has been used in direct laser cooling of molecules which is critical for increasing the number of scattered photons~\cite{Stuhl.2008}.

\section{Experimental setup}
\label{sec:Experimental}

\subsection{Ground-state molecule creation}
\label{subsec:state}

The experimental setup has been described in our previous works~\cite{Wang2013,Wang2015a,Wang2015b,Guo.16}. We prepare an optically trapped ultracold mixture of $^{23}$Na and $^{87}$Rb atom in their $\ket{F = 1, M_{F} = 1}$ state with $F$ the total angular momentum of the atom and $M_{ F}$ its projection onto the magnetic field. Weakly-bound \NaRb\, Feshbach molecules are created via magnetoassociation with the help of a Feshbach resonance at 347.64~G. A two-photon stimulated Raman process is then applied to transfer them to the lowest ro-vibrational and nuclear spin level $\ket{v'' = 0, J'' = 0, m_{J}'' = 0, m_{I}^{\rm Na} = 3/2, m_{I}^{\rm Rb} = 3/2}$ of the $X^1 {\Sigma}^+$ state. 
Here $m_{J}$ is the projection of the rotational level, $m_{I}^{\rm Na}$ and $m_{I}^{\rm Rb}$ are the nuclear spin projections of $^{23}$Na and $^{87}$Rb atoms which both have nuclear spin $I = 3/2$. 
For both the ground and the excited states, $M_{F} = m_{ J}+ m_{I}^{\rm Na}+ m_{I}^{\rm Rb}$ is always a good quantum number. 
Following this procedure, we can routinely obtain $10^4$ ground-state \NaRb\, molecules with a sample temperature of 250 nK. We note that a 335.2~G  magnetic field is always present for this work. As both electronic states have $\Omega = 0$, the transition should not be very sensitive to the small magnetic field gradient of our system.       

In the $v'' = 0$ level of \NaRb\, the $J'' = 0 \leftrightarrow J'' = 1$ transition frequency is 4.179~GHz. The rotational and nuclear spin state distribution of the ground-state UPMs can be manipulated conveniently with one- or two-photon microwave spectroscopy~\cite{Ospelkaus.10, Will.16,Gregory.16,Guo.18}. In this work, only nuclear spin states with dominating $m_{I}^{\rm Na} = m_{I}^{\rm Rb} = 3/2$ characters are used for the ground-state molecules. As will be presented in Sec.~\ref{sec:Results}, excited-state hyperfine levels with ${m}_{I}^{\rm Na}$ and ${m}_{I}^{\rm Rb}$ not equal to 3/2 can also be accessed. But the transition strength calibration is focused only on ${m}_{I}^{\rm Na} ={m}_{I}^{\rm Rb} = 3/2$ hyperfine level. For simplicity, hereafter we will omit the nuclear spin quantum numbers and label the quantum states with $\ket{J'', m_J''}$ and $\ket{J',m'_J}$ whenever possible.

\subsection{The probe laser}
\label{subsec:stab}

At zero magnetic field, the the $v'' = 0 \leftrightarrow v' = 0$ transition frequency is 338.960 THz or 884.448~nm~\cite{Docenko.2007}. In our experiment, this wavelength is provided by a home-built external cavity diode laser (probe laser) with a maximum output power of 5 mW. This laser is locked to a temperature stabilized free-running reference cavity with measured linewidth of around 1 MHz. The short-term linewidth of the probe laser is estimated to be less than 100 kHz. The long-term drift of the laser frequency, which is caused by the cavity length change during a typical measurement duration of a few hours, is on the order of one cavity linewidth (1 MHz). The cavity drift is monitored by a reference laser stabilized to another ultra-stable cavity~\cite{Guo.16,Guo.2017}. When necessary, even better long-term stability can be achieved by locking the reference cavity length to the reference laser. This is not pursued in the current work as detunings much lager 1 MHz are used for all measurements; thus the free-running cavity is sufficient. The large detuning is necessary as the transition linewidth is expected to be small and on-resonance measurement may be subject to complications from the laser linewidth.

The probe laser light is delivered to the main experiment via a single mode optical fiber. The maximum power reaching the molecule is only 0.5 mW. A beam waist of 220 $\mu$m, which is more than 10 times large than the typical sample size, is used to ensure a homogeneous illumination. The linear polarized light propagates perpendicular to the magnetic field; thus vertical (horizontal) polarized light can drive the $\pi$ ($\sigma^\pm$) transition with $\Delta M_{F} = 0~(\pm 1)$.

The frequency of the probe laser is monitored by a wavelength meter with absolute accuracy of 60 MHz. This only serves as a coarse reference for day-to-day operation. The frequency repeatability relies on the stability of the reference cavity. The relative light frequency is fine tuned with kHz resolution by a 400 MHz AOM in double-pass configuration placed before the transfer cavity.

\section{Results}
\label{sec:Results}

\subsection{Excited-state spectroscopy}
\label{subsec:HFS}

To investigate the $v' = 0$ spectrum, we prepare the ground-state \NaRb\, sample and shine on the probe laser pulse for a certain duration before detecting the remaining number of molecules. The same procedure is then repeated after stepping the light frequency. Excited-state resonances will show up as loss peaks. For the initial search, we use the highest light power and a long pulse duration to broaden the lineshape. After a loss peak is located, lower power and shorter pulse are then used for detailed frequency scans with the AOM.

\begin{figure}[t]
  \includegraphics[width=1.0\columnwidth]{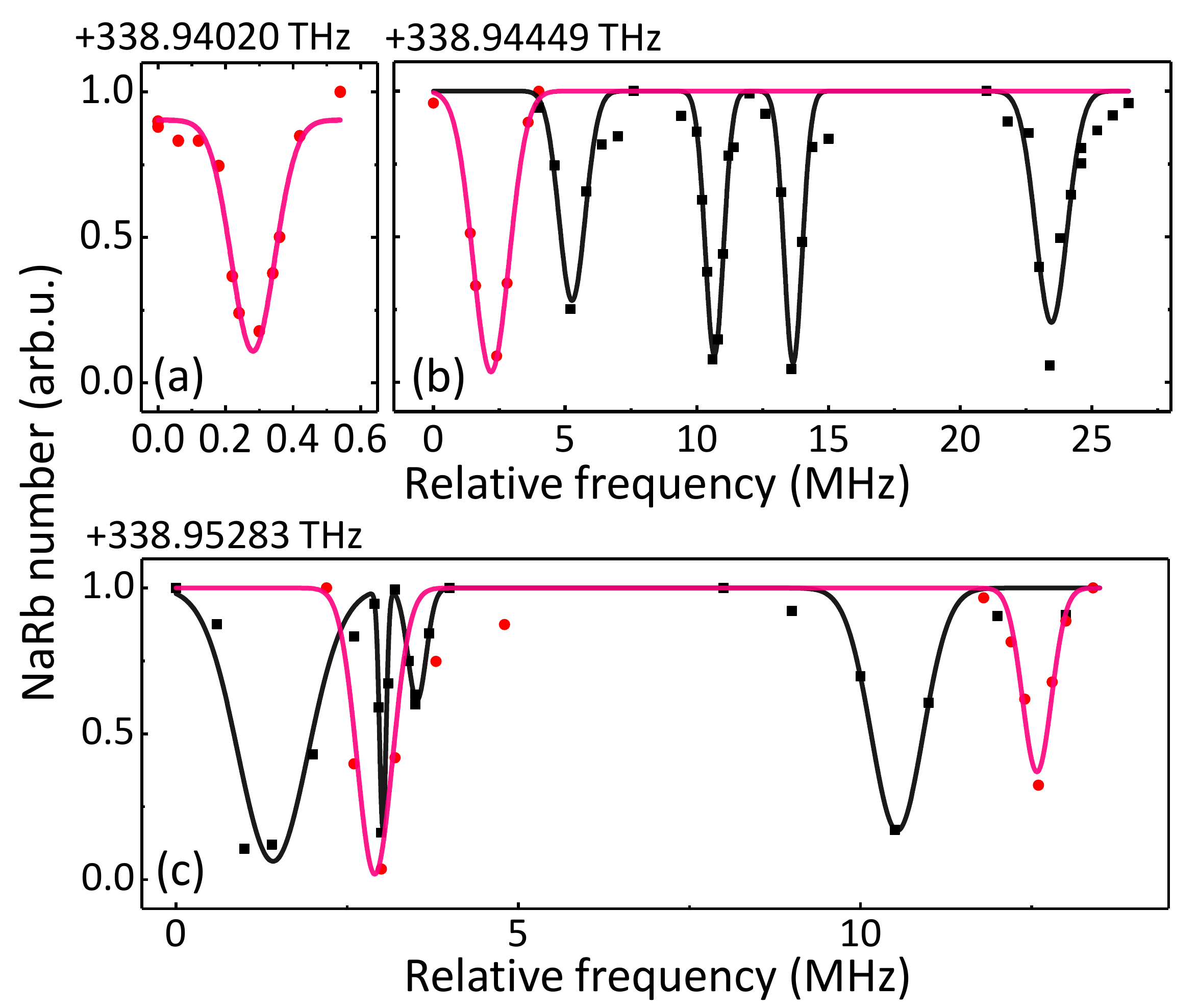}
\caption{Hyperfine structure of $J'=0$ (a), $J'=1$ (b), and $J'=2$ (c). The transition frequencies here are all relative to $\ket{J'' = 0,m_J'' = 0}$. The $R(0)$ branch spectrum in (b) is obtained with molecules in $\ket{J'' = 0, m_J'' =0}$, while the $R(1)$ and $P(1)$ branch spectra in (a) and (c) are obtained with $\ket{J'' = 1, m_J'' = 1}$ samples. Both lines probed with vertical polarized light (red circles) and horizontal polarized light are (black squares) shown. The curves are from multi-peak Gaussian fitting for extracting the line centers.}
\label{fig2}
\end{figure}

Figure~\ref{fig2}(a)-(c) show the detailed spectra of the excited state $J' = 0$, 1 and 2 levels. To access the $J'=0$ and 2 levels, the molecules are first transferred from $\ket{J'' = 0,m_J'' = 0}$ to $\ket{J = 1'',m_J'' = 0}$ with a microwave $\pi$-pulse. A common feature of the two spectra in Fig.~\ref{fig2}(b) and (c) is the many well-resolved hyperfine Zeeman structures as a result of the rotation and nuclear spin coupling and the high magnetic field~\cite{Guo.16,Guo.2017}. Due to this coupling, nuclear spins can be flipped by the electronic transitions~\cite{Guo.18}. Starting from $\ket{J'' = 0, m_J'' = 0}$ with $M_{F}'' = 3$, the $\pi$ transition can reach hyperfine structures with $M_{F}' = 3$ and the $\sigma^{\pm}$ transition can probe hyperfine structures with $M_{F}' = 2$ and 4 of the $J' = 1$ level. Similarly, starting from $\ket{J'' = 1, m_J'' = 0}$ state also with $M_{F}'' = 3$, $M_{F}' = 3$ structures can be reached by the $\pi$ transition and $M_{F}' = 2$ and 4 hyperfine structures can be reached by the $\sigma^{\pm}$ transition of the $J ' = 2$ level. Note that not all allowed $M_{F}'$ lines are observed in this work, probably due to the weaker rotation and nuclear spin couplings. 

For the $J' = 0$ level of main interest to this work, in principle several hyperfine levels can also be accessed with different light polarizations. However, as shown in Fig.~\ref{fig2}(a), we simplify the spectrum by the selection rules. Starting from $\ket{J'' = 1,m_J'' = 0,m_I^{\rm Na} = 3/2,m_I^{\rm Rb} = 3/2}$, with the vertically polarized spectroscopic light, the only level can be reached is $\ket{J' = 1,m'_J = 0,{m}_I^{\rm Na} = 3/2,{m}_I^{\rm Rb} = 3/2}$. It is important to note that besides $\ket{J ''= 1, m_J'' = 0} \leftarrow \ket{J' =0, m'_J = 0}$, spontaneous emissions to $\ket{J'' = 1, m_J'' = \pm 1}$ levels with the same nuclear spin projections are still allowed, i.e, the spontaneous emission has 3 paths in total.

For typical $\Omega = 0$ states, due to the lack of electronic angular momenta, the hyperfine structures are purely from nuclear spins. As a result, the overall span of the spectrum are often comparable to or even smaller than the nature linewidth and are thus often not resolvable~\cite{Guo.2017}. Here, the hyperfine Zeeman structures span only 30 MHz for both $J' =1$ and 2, but because of the narrow transition linewidth, they are still fully resolved. At this point, a rigorous hyperfine structure assignment, which is not the focus of this work, has not been pursued.

The experimentally measured transition frequency for the $(X^1\Sigma^+, v'' = 0, J'' = 0) \leftrightarrow (b^3\Pi, v' = 0, J' = 1)$ transition is at around 338.944 THz. This is 20 GHz lower than the calculated value at zero magnetic field. From the spectra in Fig.~\ref{fig2}, and the precisely known rotational structure of the ground state, the rotational constant of the $(b^3\Pi, v' = 0)$ level is estimated to be 2.1(1) GHz. Within the experimental uncertainty, which is dominated by the resolution of the wavelength meter, the measured rotational constant agrees well with the calculated value of 2.080 GHz.  


\subsection{Transition dipole moment}
\label{subsec:calibration}

\begin{figure}[t]
 \includegraphics[width=0.98\columnwidth]{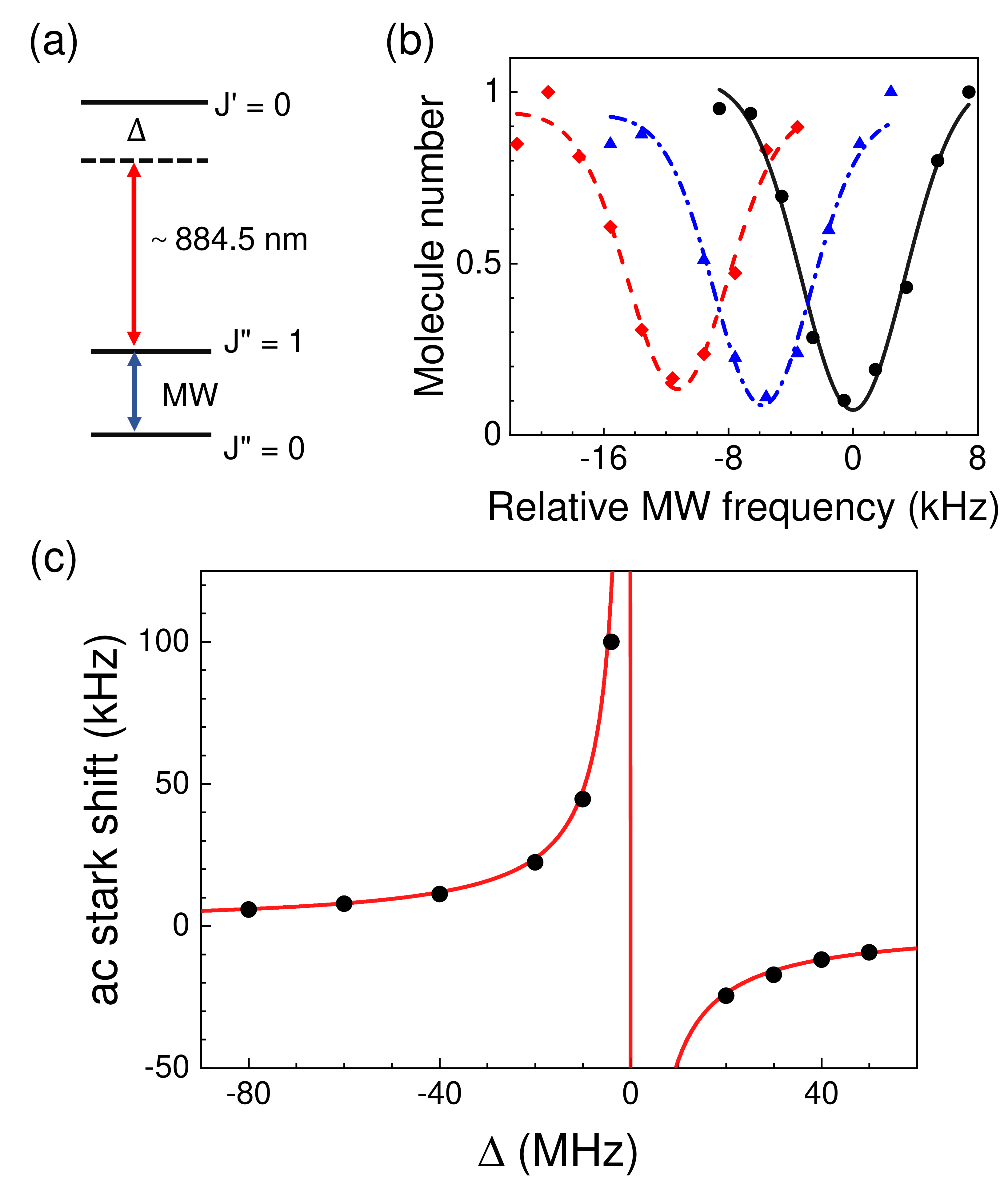}
  \caption{Measuring the $\ket{J'' =1, m_J'' = 0}\leftrightarrow \ket{J' = 0, m'_J = 0}$ TDM.  (a) ac Stark shift of $\ket{J''=1,m_J''=0}$ is induced by the probe light detuned by $\Delta$ from the $\ket{J''= 1, m_J'' = 0}\leftrightarrow \ket{J'=0, m'_J = 0}$ transition. This is revealed by scanning the MW signal driving the $\ket{J''=0,m''_J=0}\leftarrow \ket{J''=1,m''_J=0}$ transition. (b) From left to right, MW spectra for $\Delta = -40$~MHz, -80 MHz, and without the probe light. The curves are fits to Gaussian for extracting the frequency shifts. The probe light intensity is 0.23~W/cm$^2$ and the MW pulse length is 100 $\mu$s.(c) The ac Stark shift as function of $\Delta$. The solid curve is the fit for extracting the TDM $d_{P1}$. }
    \label{fig:ac stark shift}
\end{figure}

To characterize the transition strength, we measure the ac Stark shift caused by the probe light to extract the TDM $d_{P1}$ for the $\ket{J'' =1, m_J'' = 0}\leftrightarrow \ket{J' = 0, m'_J = 0}$ line. To this end, we shine the probe light to the $\ket{J'' =1, m_J'' = 0}$ sample and detect the shifted lineshape of the $ \ket{J'' =0, m_J'' = 0} \leftarrow \ket{J'' =1, m_J'' = 0}$ rotational transition with microwave (MW) spectroscopy [Fig.~\ref{fig:ac stark shift}(a)]. Several example microwave spectrum with and without the probe light are shown in Fig.~\ref{fig:ac stark shift}(b). Here the microwave pulse length is 100~$\mu$s and the typical fitting uncertainty of the line center is within $\pm 0.3$~kHz. With the probe intensity $I_0$ of 0.23 W/cm$^2$, the estimated on-resonance Rabi frequency $\omega_R$ based on the theoretical TDM should be on the level of 1~MHz. The high resolution of the microwave spectroscopy thus allows us to measure the ac Stark shifts precisely for $\Delta$ up to 100 MHz. Due to the large detuning, the ac Stark shift of the $ \ket{J'' =0, m_J'' = 0}$ level can be safely ignored. To avoid additional ac Stark shifts from the trapping light, the crossed ODT is switched off during the measurement.

Figure~\ref{fig:ac stark shift} summarizes the ac stark shifts for $\Delta$ from -10 to -80~MHz and +20 to +50~MHz. There are less data points on the $\Delta > 0$ side to avoid coupling to other hyperfine levels in the $J'' =1$ rotational state. To obtain the TDM, we first fit the data points to $-\omega_R^2/4 \Delta$, which is the ac Stark shift at the $\Delta \gg \omega_R $ limit as used here, to extract the Rabi frequency $\omega_R$. Combined with the measured probe light intensity, $d_{P1}$ is determined to be $0.0821(8)$~a.u. 

As has been discussed in~\cite{Guo.18R, He2020}, the $\ket{J'' =1, m_J'' = 0}$ hyperfine level is actually a linear combination of 82.2\% $\ket{J'' = 1,m_J'' = 0,m_I^{\rm Na} = 3/2,m_I^{\rm Rb} = 3/2}$ and 17.8\% $\ket{J'' = 1,m_J'' = 1,m_I^{\rm Na} = 1/2,m_I^{\rm Rb} = 3/2}$. The $\ket{J' =0, m_J' = 0, m_I^{\rm Na} = 3/2,m_I^{\rm Rb} = 3/2}$ excited level, on the other hand, is a spin-stretched state with no other components. As the probe light cannot flip the nuclear spin directly, we have 
\begin{equation}
\begin{split}
d_{P1}= \sqrt{0.822}\bra{v',J',m_J'} \hat{d} \sqrt{\frac{4\pi}{3}} Y_1^0 \ket{v'',J'', m_J''} \\
= 0.906 \bra{v'}\hat{d}\ket{v''}\bra{J',m_J'} \sqrt{\frac{4\pi}{3}} Y_1^0 \ket{J'', m_J''}\\
= 0.906  \bra{v'}\hat{d}\ket{v''}\sqrt{(2J'+1)(2J''+1)}\times\\ 
\begin{pmatrix}
J' & 1 & J''\\
0 & 0 & 0
\end{pmatrix}
\begin{pmatrix}
J' & 1 & J''\\
-m_J' & 0 & m_J
\end{pmatrix}
\end{split}
\end{equation} 
with $\hat{d} \sqrt{4\pi/3} Y_1^0 $ the spherical tensor component of the dipole operator corresponding to the $\pi$ transition, and $d_0 = \bra{v'}\hat{d}\ket{v''}$ the TDM between $v''$ and $v'$. Here, non-zero $d_0$ only exists between the $X^1\Sigma^+$ state and the $A^1\Sigma^+$ state coupled to the $b^3\Pi$ state. With $J'' = 1$ and $J' = 0$, the above equation can be evaluated easily to be $d_{P1}= 0.906/\sqrt{3}\times d_0$ and $d_0$ is thus 0.157(2) a.u. This value is in good agreement with the theoretical one of 0.192 a.u. This 20\% difference, which remains consistent with the overall estimated precision of the theoretical calculations, may be also due to the overestimate of the probe intensity as an result of the inaccurate focus position determination of the probe light.

\subsection{Total linewidth and transition closeness}
\label{subsec:linewidth}

The calculated linewidth of the $\ket{J'' =1, m_J'' = 0}\leftrightarrow \ket{J' = 0, m'_J = 0}$ transition with $d_{P1}$ is only $3.13(6)$ kHz. As $m_J' = 0$ can also decay to $m_J'' = \pm1$ with equal branching ratios as that to $m_J'' = 0$, the partial linewdith of the $J''=1 \leftarrow J' =0$ transition should be $\Gamma_{\rm partial} = 9.4(2)$~kHz. Apparently, the linewidth of the measured lineshape in Fig.~\ref{fig2}(a) is much larger than this which indicates the existence of other possible spontaneous decay channels.

\begin{figure}[t]
  \includegraphics[width=1\columnwidth]{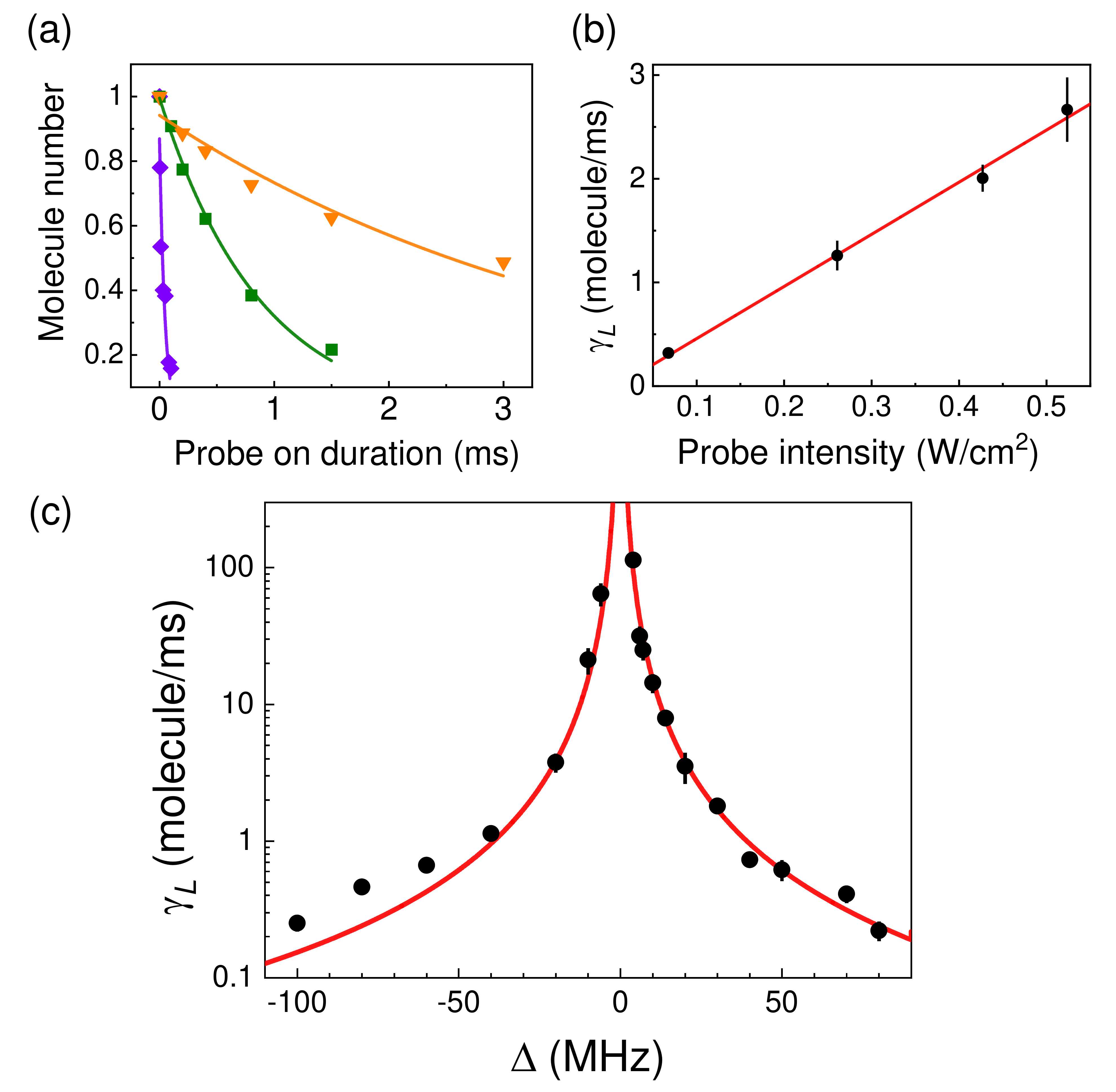}
\caption{Measurement of the total linewidth $\Gamma$ of $v''=0$. (a) Loss of molecules from the optical lattices for probe light detunings $\Delta$ at -10 MHz (purple diamonds), -40 MHz (green squares), and -100 MHz (orange triangles). The solid lines are exponential decay fitting for obtaining the loss rate $\gamma_L$. (b) At a fixed $\Delta$, $\gamma_L$ has a linear dependence on the probe intensity. (c) $\gamma_L$ as function of $\Delta$ at a probe light intensity of $I_0 = 0.52$~W/cm$^2$. The red solid curve is the fit with Eq.~\ref{eq2} from which $\Gamma$ is determined to be 225(5)~kHz. }
\label{fig:loss rate}
\end{figure}

As the FCFs from $(b^3\Pi, v' = 0)$ to $(X^1\Sigma^+, v'' \neq 0)$ levels are very small, spontaneous decay to these levels cannot be the main cause of the larger measured linewidth. As shown in Fig.~\ref{fig1}, the only other possible decay path is to the dissociation continuum of the lowest triplet state $a^3\Sigma^+$. Taking this decay path into account, the calculated Einstein coefficients, which include both the FCF and the TDM shows that the spontaneous emission has 75\% probability to the $(X^1\Sigma^+, v'' = 0)$ level, and 77.7\% probability to all \Xstate\ levels. The spontaneous emission probability to $a^3\Sigma^+$ is 22.3\%. The calculated radiative lifetime of the $(b^3\Pi, v' = 0)$ state is 6.97~$\mu$s, which corresponds to a natural linewidth of 22.8 kHz, in reasonable agreement with $\Gamma_{\rm partial}$ above.

Limited by the probe laser linewidth, extracting the total linewidth of the excited state using the loss spectrum~\cite{Guo.2017} in Fig.~\ref{fig2}(a) may not be reliable. To obtain the total linewidth of the excited state, instead we measure the off-resonance photon scattering caused by the probe light. To mitigate the two-body collisional loss, the molecules are loaded into 1064 nm 3-D optical lattices to isolate them from each other. At a lattice depth of about 50 recoil energy of \NaRb, the sample has a trap lifetime of over 6 seconds~\cite{note1} which is much longer than the time needed for the photon scattering measurement. As the branching ratios to $J'' =0$ and $\pm 1$ are identical, and there are also other hyperfine levels as possible allowed decay channels, the probability of decaying back to $\ket{J'' = 1, m_J'' = 0}$ is no more than 25\%, even without considering the leakage to $a^3\Sigma^+$ state. Therefore, each molecule can only scatter at most 1.3 photons before getting lost to the dark states. For simplicity, we assume the photon scattering rate to be the same as the molecule loss rate.

Figure~\ref{fig:loss rate}(a) shows three example loss curves of $\ket{J'' = 1, m_J'' = 0}$ molecules from the 3-D lattices in presence of the probe light at different $\Delta$. Without the probe light, the number of molecules is totally flat within the measurement time. Fit the loss curves to the one-body exponential decay, the loss rate $\gamma_L$ for each $\Delta$ is extracted and plotted in Fig.~\ref{fig:loss rate}(c). We fit the measured $\gamma_L$ with~\cite{Bause.2019} 
\begin{equation}
\gamma_L = \frac{3 \pi c^2}{2 \hbar \omega^3} \frac{\Gamma_{\rm{partial}} \Gamma}{\Delta^2} I_0
\label{eq2}
\end{equation}
with $\omega$ the transition frequency, and the probe intensity at $I_0 = 0.52~\rm{W/cm^2}$. The total linewidth of the $\ket{J' = 0, m_J' =0}$ level obtained from the fitting is $\Gamma = 225(5)$ kHz, which is significantly larger than the calculated value of 22.8 kHz.


\section{Discussion}
\label{sec:Discussion}

The much larger value of $\Gamma$ compared to $\Gamma_{\rm partial}$ proves that the spontaneous emission decay does not dominate the lifetime of the ($b^3\Pi, v' = 0$) level. Looking at the PECs derived from spectroscopic analysis in Fig.~\ref{fig1}, it is expected that the repulsive branch of the $a^3\Sigma^+$ curve comes close to the bottom of the $b^3\Pi$ curve. This trend is also confirmed by our theoretical calculations. Thus rotational predissociation of the ($b^3\Pi, v' = 0$) level toward the $a^3\Sigma^+$ dissociation continuum would be the main decay process. Both experimental and theoretical PECs are not precise enough -especially the $a^3\Sigma^+$ repulsive branch- to yield a definite interpretation. On the other hand, the present spectroscopic analysis could provide constraints on that curve in the framework of a dynamical model for rotational predissociation. This will be investigated in future work.


It is thus very difficult, if not possible, to plug the leakage for a more closed $\ket{J'' = 1, m_J'' = 0}\leftrightarrow \ket{J'=0, m_J ' =0}$ transition. Practically speaking, this transition is thus not very useful for direct laser cooling of ground-state \NaRb. On the other hand, as shown in~\cite{guan20}, this transition may still be used for non-destruction dispersive imaging of high optical depth \NaRb\, samples. Finally, one might think that the $\Gamma$ value obtained here indicates that photon scattering induced loss may be a severe issue for the optical shielding scheme based on this transition~\cite{xie20}: this constraint could be significantly relaxed considering only $\Gamma_{\rm partial}$, as suggested by the interpretation above. 

\begin{figure}[th]
  \includegraphics[width=0.8\columnwidth]{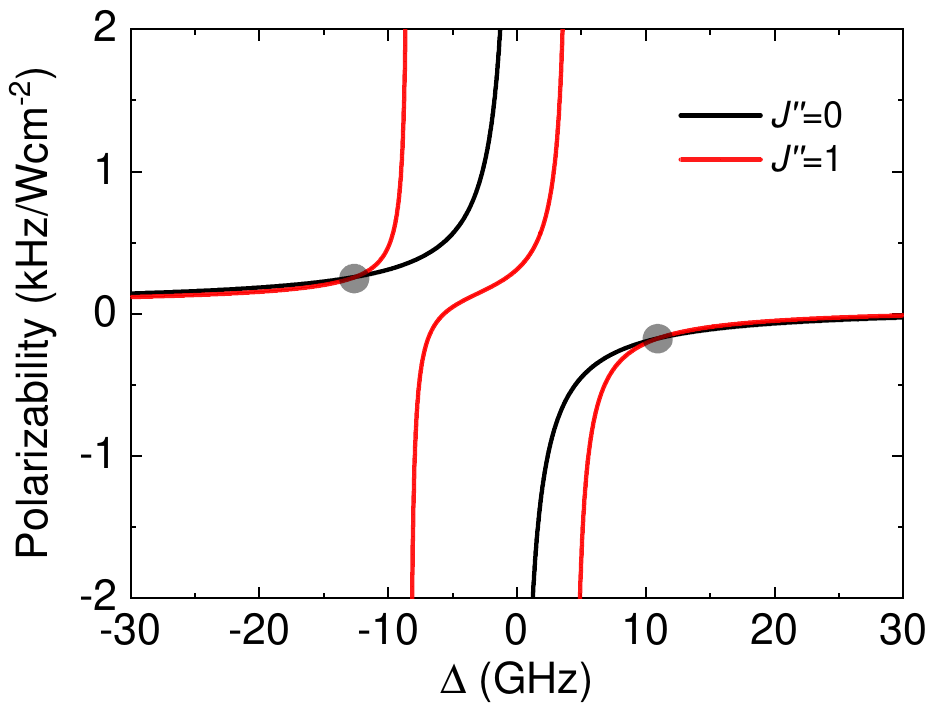}
\caption{The calculated polarizabilities of $J''=0$ (black line) and $J'' = 1$ (red line) molecules. The detuning $\Delta$ is relative to the $\ket{J''=0,m''_J=0} \leftrightarrow \ket{J'=1,m'_J=0}$ transition. The light is linear polarized with a $60^\circ$ polarization angle relative to the quantization axis defined by the magnetic field. For this polarization angle, there are two magic $\Delta$ (gray dots) where the polarizabilities cross each other.}
\label{fig:pol}
\end{figure}

Another application of this transition, as demonstrated in~\cite{Bause.2019}, is to create a ``magic'' trap so that $J''=0$ and 1 experience exactly the same trapping potentials. This will be an important way to optimize the rotational coherence necessary for future applications in quantum computing with ultracold polar molecules~\cite{ni.18,Sawant2020}. Previously, we already studied this kind of magic trap for \NaRb\, with 1064 nm light. Using the anisotropic polarizability of $J'' = 1$, it is always possible to find a polarization angle such that the rotational coherence is the longest. However, due to the intensity-dependent coupling between hyperfine levels, the potential matching between $J''=0$ and 1 are not perfect~\cite{Lin2021}. For the trap formed by 884.5 nm light, besides the background polarizability from all the other vibrational levels, the near resonance polarizability from the narrow linewidth $v'' =0 \leftrightarrow v' =0$ transition also has a large contribution. The overall potential can thus be fine tuned by both the polarization angle and the near resonance detuning. As a result, a near perfect potential matching becomes possible~\cite{Bause.2019}. 

Figure.~\ref{fig:pol} shows the calculated polarizabilities of $J''=0$ and 1 versus $\Delta$ with a polarization angle of $60^\circ$. Here $\Delta$ is relative to the $\ket{J'' = 0, m_J'' =0}\leftrightarrow \ket{J' = 1, m_J'=0}$ resonance. The background polarizabilities parallel and perpendicular to the molecular axis at 884.5 nm used in this calculation, $\alpha_{\parallel}=126~h\times$ Hz/(W/cm$^2$) and $\alpha_{\perp}=24~h\times$ Hz/(W/cm$^2$), respectively, are obtained following the general formula given in Ref.~\cite{Vexiau17}. The polarizability calculation procedure is similar to that for $^{23}$Na$^{40}$K~\cite{Bause.2019}, so will not be described in detail here. Two crossings, one at $\Delta = -12.5$~GHz, the other at $\Delta = 10.9$~ GHz can be observed from the plot. At $\Delta$ = -12.5 GHz, a focused laser beam with intensity of 0.4 kW/cm$^2$ can generate a trapping potential with 2~$\mu$K depth. Under this condition, the trap lifetime limited by photon scattering is estimated to be near 130 ms.

\section{Conclusion}
\label{sec:Conclusion}

So far, the electronic spin forbidden $(X^1\Sigma^+, v'' = 0) \leftrightarrow (b^3\Pi, v' = 0)$ transition has been investigated in ultracold KRb~\cite{Kobayashi.14}, $^{23}$Na$^{40}$K~\cite{Bause.2019}, and now \NaRb\, molecules. While the FCF for this transition is almost diagonal for all species, the excited-state linewidth is determined by the competition between (i) the balance of the spontaneous decay between the $X^1\Sigma^+$ and $a^3\Sigma^+$ states, largely influenced by the relevant transition dipole moments, and (ii) by the possibility for rotational predissociation. The TDM of the $(a^3\Sigma^+ \leftrightarrow (b^3\Pi, v' = 0)$ transition shows a strong variation from one species to another. For \NaRb, this TDM is large which results in a much larger total linewidth of the $(b^3\Pi, v' = 0)$ level than in both KRb~\cite{Kobayashi.14} and $^{23}$Na$^{40}$K~\cite{Bause.2019}. Nevertheless, the measured linewidth here is still about 100 times smaller than the 20 MHz linewidth of typical allowed molecular transitions. It is thus still valuable for engineering and detecting of trapped ultracold \NaRb\, molecules. 

This special transition can also be found in other heteronuclear bi-alkali ground-state molecules. For those cases, although experimental investigations are still not available, our study shows that reasonably accurate prediction of both the partial and total linewidths can be obtained from theoretical calculations.  This will be presented in a forthcoming work.

\section{Acknowledgments}

We thank Xin Ye for his contributions at the early stage of this project. This work was supported by the Hong Kong RGC General Research Fund (grants 14301818, 14301119, and 14303317) and the Collaborative Research Fund C6005-17G and by the French \textit{Agence Nationale de la Recherche} (ANR), under the joint ANR-RGC agreement ANR-13-IS04-0004-01.

\bibliography{ref}

\end{document}